\documentclass[%
reprint,
superscriptaddress,
 amsmath,
 amssymb,
 aps,
 prl
]{revtex4-2}

\usepackage{graphicx}
\usepackage{subfig}
\usepackage{dcolumn}
\usepackage{bm}
\usepackage{caption}
\usepackage{balance}
\usepackage{hyperref}
\hypersetup{
    colorlinks=true,
    linkcolor=red,
    filecolor=red,
    urlcolor=red,
    citecolor=blue, 
}
\usepackage[font=small,labelfont=bf]{caption}
\usepackage[symbol]{footmisc}

\newcommand{\bra}[1]{\left\langle #1 \right|}
\newcommand{\ket}[1]{\left|#1\right\rangle}

\begin{document}

\title{Achieving Heisenberg Scaling on Measurement of a Three-Qubit System\\via Quantum Error Correction}
\author{\small Le Hu}
\email{lhu9@ur.rochester.edu}
\affiliation{Department of Physics and Astronomy, University of Rochester, Rochester, New York 14627, USA}
\affiliation{Institute for Quantum Studies, Chapman University, 1 University Drive, Orange, CA 92866, USA}
\author{\small Shengshi Pang}
\affiliation{School of Physics, Sun Yat-Sen University, Guangzhou, Guangdong 510275, China}
\author{\small and Andrew N. Jordan}
\affiliation{Institute for Quantum Studies, Chapman University, 1 University Drive, Orange, CA 92866, USA}
\affiliation{Department of Physics and Astronomy, University of Rochester, Rochester, New York 14627, USA}

\date{\today}

\begin{abstract}
In many-body quantum systems, the quantum Fisher information an observer can obtain is susceptible to decoherence. Consequently, the Heisenberg scaling in quantum enhanced metrology cannot usually be achieved. We show, via two different approaches and under certain approximations, that by applying periodic quantum error corrections, we can achieve the Heisenberg scaling both for time (\textit{i.e.} QFI $\propto t^2$) and the number of atoms $s$ (\textit{i.e.} QFI $\propto s^2$) for an extended period of time on measurement of detuning frequency of the Tavis-Cummings model, where initially maximally entangled two-level atoms interact with a single cavity mode.
\end{abstract}
\maketitle



\section{\label{sec:level1}Introduction}

One of the fundamental topics in metrology is how to improve the precision of measurements. Quantum metrology, which has a growing number of applications in modern physics \cite{giovannetti2006quantum,giovannetti2011advances,schnabel2010quantum,taylor2016quantum,nolan2017quantum,valencia2004distant,albarelli2020perspective}, is not an exception. In quantum metrology, quantum-related characteristics such as quantum entanglement and quantum squeezing are utilized to improve the precision of measurements \cite{jordan2019time,giovannetti2004quantum,huelga1997improvement,holland1993interferometric}, an enhancement that classical metrology does not enjoy. However, as any quantum system to be studied cannot be perfectly isolated from the environment, such enhancement can be easily undermined by decoherence \cite{shaji2007qubit,alipour2014quantum}. More explicitly, in many-atom systems, the Heisenberg scaling says that the quantum Fisher information (QFI), a measure of precision in measurements, can scale as $N^2$ with good quantum correlation between systems, but only as $N$ without correlation \cite{giovannetti2004quantum}, where $N$ is the number of systems. A similar law also applies in terms of time, which means that the QFI scales as $T^2$ for systems with good coherence but only as $T$ for decoherent ones, where $T$ is the time at measurement.

\begin{figure}[!hp]
\centering
\includegraphics[width=0.387\textwidth]{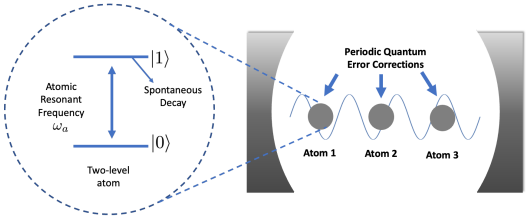}
\captionsetup{justification=centering}
\caption{
\label{fig:epsart1} Three two-level atoms in a cavity with periodic QECs.} 
\end{figure}

Many attempts \cite{pang2017optimal,zhou2018achieving,chen2020fluctuation,demkowicz2014using}, thereby, have been made to preserve the coherence of the system, so that the Heisenberg scaling can be achieved for a longer period of time for the purpose of maximizing the QFI. Among them, the idea of quantum error correction (QEC) \cite{shor1995scheme, preskill1998reliable, kitaev1997quantum, knill1998resilient, gottesman1998theory, gottesman2010introduction, kessler2014quantum, arrad2014increasing, ozeri2013heisenberg, lu2015robust} has been increasingly popular in recent years, which can be very useful when applied in quantum enhanced metrology \cite{dur2014improved}.

In this paper, by making use of periodic QEC, we show via two different approaches how the Heisenberg scaling is achieved both for the time scaling (\textit{i.e.} QFI $\propto t^2$) and the number of atoms $s$ scaling (\textit{i.e.} QFI $\propto s^2$), for the measurement on the detuning frequency of the Tavis-Cummings model. We first study the three-qubit case and then discuss the generalization to the arbitrarily many qubits case.
The results are obtained under the approximation that the number of photons in the cavity is sufficiently large, which makes the analytical calculation of the evolution of the system simpler. We show that the method works well in both cases where the photons are in the Fock state and in the coherent state. While a large number of photons in a Fock state is too ideal for a real environement, the idea of such an approximation is that if we treat photons, regardless of them being in the Fock state or in the coherent state, as noise that undermines the initial coherence of the atomic system, and then if the QEC works well in a noisy environment, it is reasonable to expect the QEC will work as well, if not better, in a less noisy environment, such as in a vacuum.

\section{FORMULATION OF THE PROBLEM}
Let us first consider the Tavis-Cummings model \cite{tavis1968exact,tavis1969approximate,agarwal2012tavis} for the three-atom case in the region where the rotating-wave approximation works.  Those two-level atoms interacts with a single cavity mode, experimentally realizable by superconducting cavity (Fig.~\ref{fig:epsart1}), which usually has a very high quality factor. The Hamiltonian is given by
\begin{equation}\label{eq1}
\hat{H}=\hbar \omega_{c} \hat{a}^{\dagger} \hat{a}+\hbar \omega_{a} \sum_{i=1}^{3} \frac{\hat{\sigma}_{z}^{(i)}}{2}+\frac{\hbar \Omega}{2} \sum_{i=1}^{3}\left(\hat{a} \hat{\sigma}_{+}^{(i)}+\hat{a}^{\dagger} \hat{\sigma}_{-}^{(i)} \right)	
\end{equation}
where $\hat{a}$ and $\hat{a}^\dagger$ are the annihilation and creation operator for the cavity mode, $\hat{\sigma}_z^{(i)}$ is the Pauli-$z$ matrix of the $i$-th atom, $\omega_c$ is the angular frequency of the field, and $\omega_a$ is the atomic transition frequency. The raising operator $\hat{\sigma}_+^{(i)}$ and lowering operator $\hat{\sigma}_-^{(i)}$ of the $i$-th atom is defined as $\sigma_\pm^{(i)}=\sigma_x^{(i)} \pm i \sigma_y^{(i)}$.

To obtain the maximum QFI of the frequency, it is sufficient to let the atoms initially be at maximally entangled state (Greenberger-Horne-Zeilinger state) \cite{greenberger1989going,giovannetti2006quantum}
$
\ket{\Psi(0)}=\frac{1}{\sqrt{2}}(\ket{000}+\ket{111}).
$
Here we consider the case where the environment is initially in the Fock state $\ket{n}$ (for the analysis for the coherent state, see Method II). The density matrix of the whole system at time $t=0$ can then be described as
	$\rho(0) = \rho_S(0)\otimes\rho_E(0)$,
where 
	$\rho_S(0) = \ket{\Psi{(0)}}\bra{\Psi(0)}$ and
	$\rho_E(0) = \ket{n}\bra{n}$, which at time $t$ evolves to $\rho(t)=U(t)\rho(0)U^\dagger(t),$
where $U(t) = \exp(-iHt/\hbar)$. Suppose we then make local measurements on the atoms, meaning that all the information we can possibly gain is contained in the partially traced density matrix $\rho_S(t)=\operatorname{Tr}_E(\rho(t) )$, which shall be used to calculate the QFI $\mathcal{F}_{\tau}$. This can be done by utilizing \cite{knysh2011scaling,liu2013phase,liu2019quantum}
\begin{equation}\label{eq7}
\begin{aligned} \mathcal{F}_{\tau}=& \sum_{i=1}^{n} {\lambda_{i}^{-1}}(\partial_\tau \lambda_i)^{2}+4 \sum_{i=1}^{n} \lambda_{i}(\left\langle\partial_\tau v_i|\partial_\tau v_i\right\rangle-\left|\left\langle v_{i}|\partial_\tau v_i\right\rangle\right|^{2}) \\ &-8 \sum_{i \neq j}^{n} \frac{\lambda_{i} \lambda_{j}}{\lambda_{i}+\lambda_{j}}\left|\left\langle v_{i}|\partial_\tau v_j\right\rangle\right|^{2} \end{aligned}	
\end{equation}
where $\lambda_i$ and $\ket{v_i}$ are the eigenvalue and the eigenstate of $\rho_S(t)$, respectively, and $\tau$ (which in our case is the detuning frequency), contained in $\rho_S(t)$, is the parameter to be measured. The QFI obtained directly through this way, \textit{i.e.} without any QEC, achieves the Heisenberg scaling only at the time very shortly after the system begins to evolve, when the decoherence has not yet done much to the system. As time elapses, the atoms start to lose coherence quickly as they become entangled with the cavity, which is manifested in our cases as bit-flipping as can be seen in  $\rho_S(t)$. This effect will lower the maximum QFI.

The idea then is, to apply a QEC that un-flips wrongly flipped qubits, at a short time $\epsilon$ after the systems begin to evolve when the coherence of the system is still mostly preserved. Note that we assume the QEC happens fast enough such that at most one qubit is wrongly flipped. The corrected system then begins to evolve and lose coherence again and we must reapply the QEC, e.g. at the fixed time interval $\epsilon$. This should save the system from decoherence for an extended period of time until remedies eventually fail. It is naturally expected that the shorter the time interval $\epsilon$ is, the better the result is, as less coherence is lost every round. This means that in principle the coherence can be preserved well for a long time $T$ as long as the time interval $\epsilon$ is small enough.

\section{Approximate Solution -- Method I}
\subsection{Derivation}
While the appealing idea above is not complicated, computing it for many-atom cases is non-trivial, especially if the number of atoms $s$ becomes large. The reason is that the dimension of the Hilbert space grows by $2^s$, which along with the noncommutativity between $\hat{a}$ and $\hat{a}^\dagger$ makes its diagonlization difficult.

To get around this, we take the approximation $n \gg s$ such that
\begin{equation} \label{eq8}
\begin{aligned} 
	&(\hat{a})^i\ket{n+i} = \sqrt{(n+i)!/n!} \ket{n} \approx \sqrt{n^i}\ket{n}\\
	&(\hat{a}^\dagger)^i\ket{n} = \sqrt{(n+i)!/n!} \ket{n+i} \approx \sqrt{n^i}\ket{n+i},
\end{aligned}
\end{equation}
where $i = 1, 2, \cdots, s$. This approximation essentially says that $([\hat{a},\hat{a}^\dagger] \ket{n})/(\bra{n}\hat{a}^\dagger \hat{a}\ket{n}) \approx 0$ for large enough number states $\ket{n}$; \textit{i.e.} the noncommutativity between $\hat{a}$ and $\hat{a}^\dagger$ affects very little on the eigenvalues when $n \gg s$. 
To begin with the three-atom case, initially $\rho(0)$ is given by $\rho_{11}=\rho_{18}=\rho_{81}=\rho_{88}=1/2\rho_E(0)$ and all other entries are zero. After the state evolves for a certain time, we then apply the QEC map 
\begin{equation}
\mathcal{E}_{\text {corr }}(\rho)=P_{0} \rho P_{0}+\sum_{i=1}^{3} X_{i} P_{i} \rho P_{i} X_{i}	,
\end{equation}
\begin{equation}
\begin{aligned}
P_{0}&=|000\rangle\langle 000|+| 111\rangle\langle 111| \\
P_{1}&=|100\rangle\langle 100|+| 011\rangle\langle 011|\\
P_{2}&=|010\rangle\langle 010|+| 101\rangle\langle 101|\\
P_{3}&=|001\rangle\langle 001|+| 110\rangle\langle 110|,
\end{aligned}
\end{equation}
where $X_{i}$ is the Pauli-$X$ gate acting on the $i$-th qubit. It is straightforward to show that no matter what the explicit expression of $\rho(t)$ is, with QEC applied, the resultant density matrix $\mathcal{E}_{\text {corr }}(\rho(t))$ contains up to four non-zero terms, which reside at the four corners of the density matrix. The resultant matrix can then be transformed into a block-diagonal one, where only the $4 \times 4$  block matrix at the lower right is nonzero: 

\begin{equation}
	\begin{pmatrix}
	p_{11} & &p_{18}\\
	&0_{6\times6}\\
	p_{81} & & p_{88}
\end{pmatrix} \rightarrow
\begin{pmatrix}
	0_{4\times4} &&\\
	&p_{11}&&p_{18}\\
&&0_{2\times2}&\\
	&p_{81}&&p_{88}\\
\end{pmatrix}.
\end{equation}
The $4 \times 4$ submatrix has bases $\ket{-\frac{3}{2}}_a, \ket{-\frac{1}{2}}_a, \ket{\frac{1}{2}}_a$, and $ \ket{\frac{3}{2}}_a$, where $\ket{i}_a$ denotes there are $(i+\frac{3}{2})$ atoms in excitation. This collective-spin treatment neglects the detailed configuration of the atoms, but instead consider only the number of atoms being excited, which helps to switch the dimension of the Hilbert space of the atoms from $2^3$ to 4. The original Hamiltonian can be correspondingly rewritten as 
\begin{equation}
\begin{aligned}
H &= \underbrace{\hbar \omega_{c}\left(\hat{a}^{\dagger} \hat{a}+\frac{\hat{S}_{z}}{2}\right)}_{H_\text{I}} + \underbrace{\hbar \delta \frac{\hat{S}_{z}}{2}+\frac{\hbar \Omega}{2}\left(\hat{a} \hat{S}_{+}+\hat{a}^{\dagger} \hat{S}_{-}\right)}_{H_\text{II}}
\end{aligned}
\end{equation}
where $\delta=\omega_a-\omega_c$ is the detuning frequency, and $\hat{S}_{\alpha}=\sum_{j=1}^{3} \hat{\sigma}_{\alpha}^{(j)}$ where $\alpha \in \{z,+,-\}$. Note that the dimension of $\hat{S}_{\alpha}$ is 4, the same as the $4 \times 4$ sub-matrix we introduced above. Since $[H_{\text{I}}, H_{\text{II}}] =0$, we have
$
U(t)=U_\text{I}(t) U_\text{II}(t).
$
Then at time $t=\epsilon$,
\begin{equation}\label{eq22}
\begin{aligned} \rho(\epsilon) &=U_\text{II}(\epsilon) U_\text{I}(\epsilon) \rho(0) U_\text{I}^{\dagger}(\epsilon) U_\text{II}^{\dagger}(\epsilon)
\end{aligned}	
\end{equation}
The $U_\text{II}(\epsilon)$ can be computed as an ordinary matrix exponential as we have considered the approximation case where $n \gg 3$ such that $[a, a^\dagger] \approx 0$ [Eq.\,(\ref{eq8})]. 
Tracing over the environment by $\rho_S(\epsilon) = \sum_m \bra{m} \rho(\epsilon) \ket{m}$, where $\{ \ket{m}\}$ is the basis in the Fock space, the only non-zero entries are given by
\begin{equation}\label{eq25}
\begin{aligned}
{\rho_S}_{11}(\epsilon)&={\rho_S}_{44}(\epsilon)=\chi^3(\epsilon)+(1-\chi(\epsilon))^3,\\
{\rho_S}_{22}(\epsilon)&={\rho_S}_{33}(\epsilon)=	3\chi^2(\epsilon)(1-\chi(\epsilon))+3 \chi(\epsilon)(1-\chi(\epsilon))^{2}\\
{\rho_S}_{14}(\epsilon)&={\rho^*_S}_{41}(\epsilon)=p_{14}(\epsilon)e^{i3\omega_c\epsilon}\\
\end{aligned}
\end{equation}
where
\begin{equation}
\begin{aligned}
	\chi(\epsilon)&=\frac{(n+1) \Omega^{2} \sin \left(\frac{1}{2} \Delta \epsilon \right)}{\Delta^{2}}\\
\Delta&=\sqrt{\delta^{2}+(n+1) \Omega^{2}},\\
\end{aligned}
\end{equation}
and
\begin{equation}
\begin{aligned}
p_{14}(\epsilon)=\frac{1}{16\Delta^4}\left[ i \delta \left( 3 \chi(\epsilon)+\left( \frac{4\delta^2}{(n+1)\Omega^2}+3\right)\chi(3 \epsilon)\right)+ \right.\\
\left. 3 \chi(\epsilon) \Delta \cot\left(\frac{1}{2}\Delta \epsilon\right)+\frac{ \left( \Delta^2+3\delta^2\right)}{\Delta}\cos{\left(\frac{3}{2}\Delta \epsilon\right)}\right]^2.
\end{aligned}
\end{equation}

Note how the diagonal terms are ``symmetric", which means if we apply QEC \textit{before} tracing over the environment \textit{and then} do the tracing over part, the result should be
\begin{equation}
\rho_S^\prime(\epsilon)=\sum_m\bra{m}\mathcal{E}_{\text {corr }}(\rho(\epsilon))\ket{m},
\end{equation}
of which the only non-zero entries are
\begin{equation} \label{neweq1}
{\rho_S^\prime}_{14}(\epsilon)=\frac{1}{2}e^{i3\omega_c \epsilon}\sum_m \bra{m}(U_\text{II}(\epsilon))_{11} \rho_E(0) (U_\text{II}^\dagger(\epsilon))_{44}	\ket{m},
\end{equation}
${\rho_S^\prime}_{41}(\epsilon)={\rho_S^\prime}_{14}^*(\epsilon)$ and ${\rho_S^\prime}_{11}(\epsilon)={\rho_S^\prime}_{44}(\epsilon)=1/2$. Note that in Eq. ~(\ref{neweq1}) terms that did not survive after partial tracing have been omitted. The prime on $\rho_S^\prime(\epsilon)$ denotes that the QEC has been applied. Notice the similarity between $\rho_S^\prime(\epsilon)$ and $\rho_S(0)$; the only difference is the off-diagonal corner entries $\rho_{14}$ and $\rho_{41}$, the evolution of which indicates the loss of coherence despite QEC. 
\begin{figure*}[!htp]

\subfloat{\includegraphics[width=.32\textwidth,scale=1]{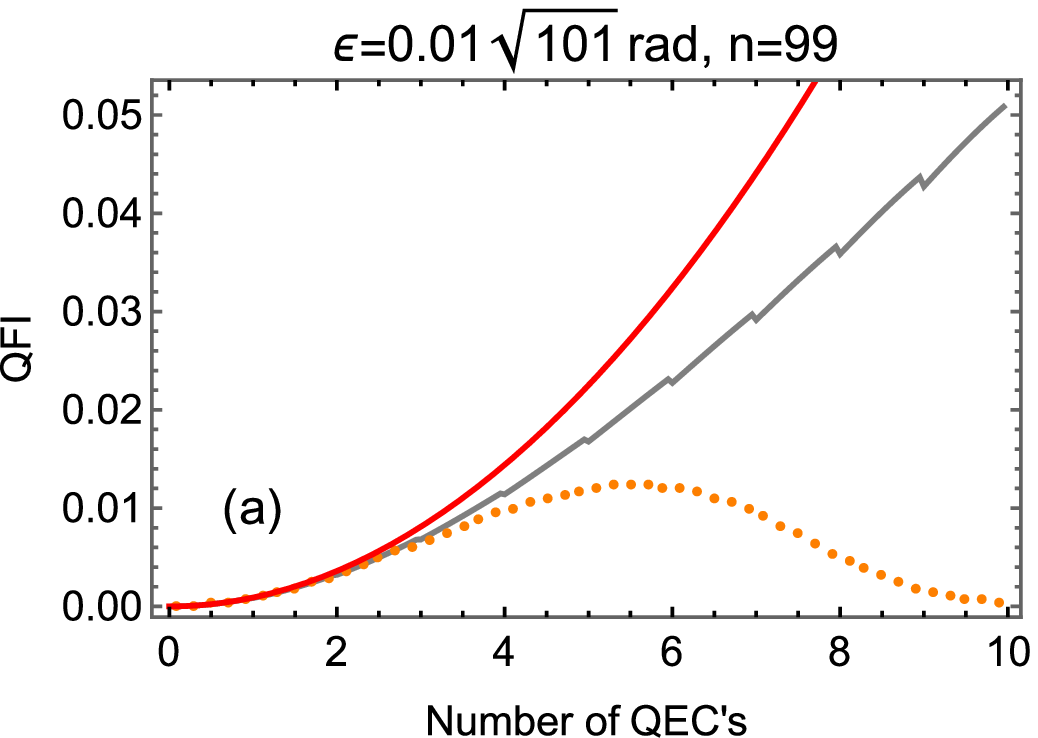}\hfill}
\subfloat{
\includegraphics[width=.325\textwidth,scale=1]{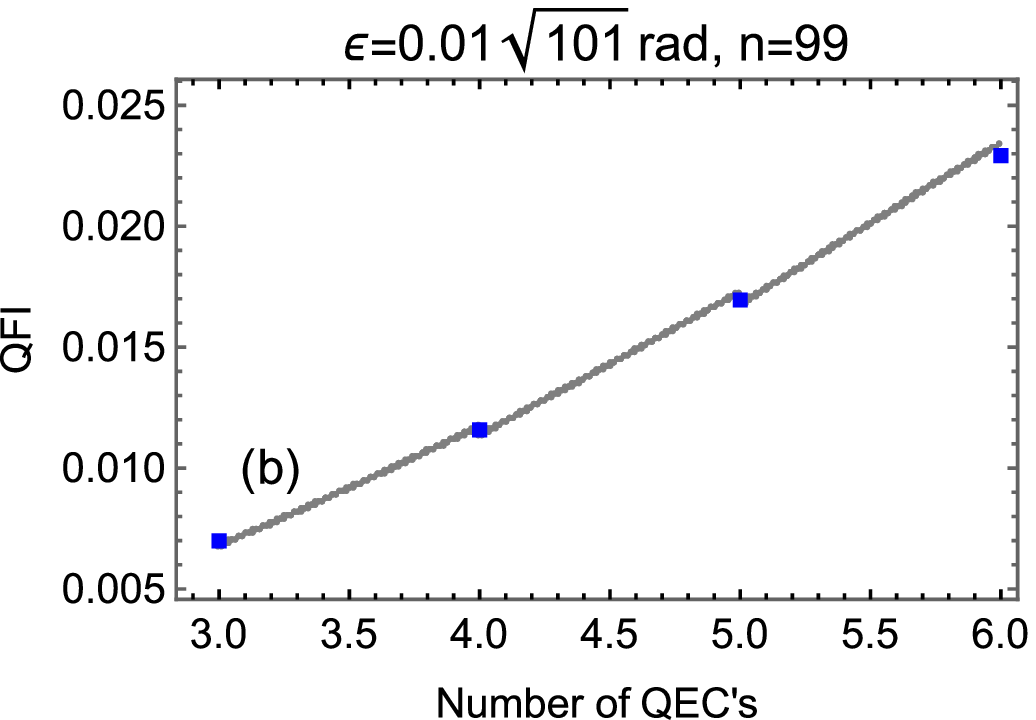}\hfill}
\subfloat{
\includegraphics[width=.33\textwidth,scale=1]{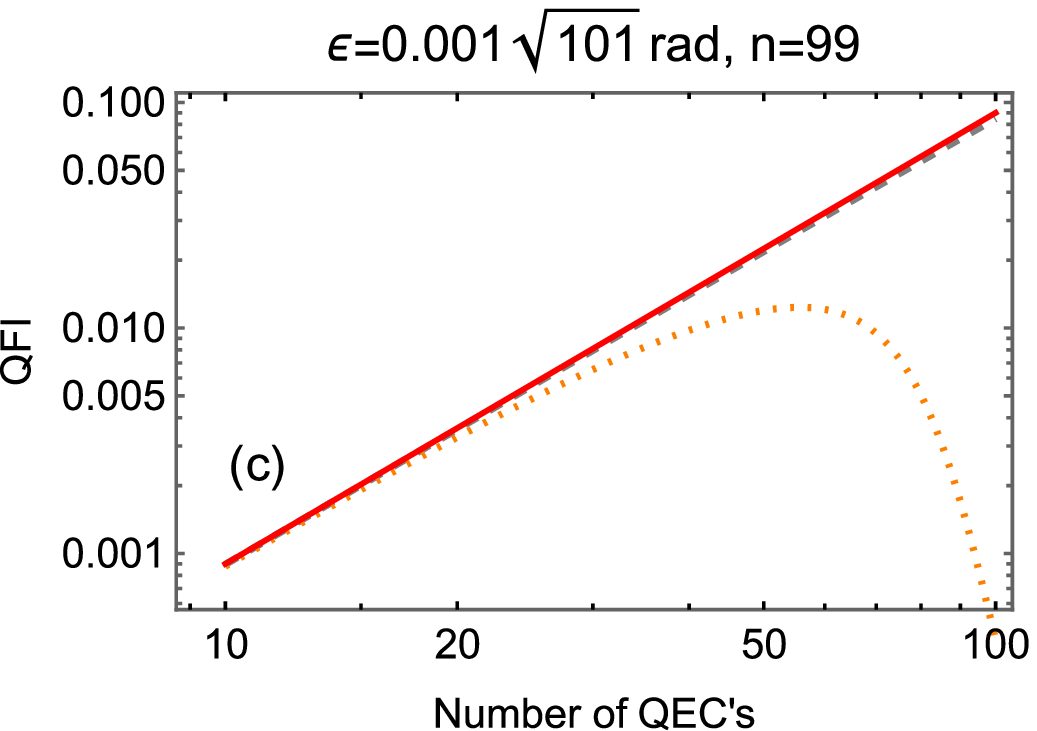}}
\caption{\label{fig:epsart2} The QFI regarding the detuning frequency $\delta$ obtained at time $t$, using the approximation method I, for the uncorrected case (orange dotted curve), the QEC case [blue square in panel (b)], the case in between two successive QEC [gray zigzag line in panel (a), straight line in panel (b) and dashed line in panel (c)], and the ideal Heisenberg scaling case $\mathcal{F}_{\delta}(t)=9t^2$ (red curve). The time interval $\epsilon = 1/2 \Delta t_0$ is dimensionless where $\Delta = \sqrt{\delta^2+(n+1)\Omega^2}$ and $t_0$ is the dimensional time. We take $\delta=2$, $\Omega=2$, $n=99$, and $\rho_E=\ket{n}\bra{n}$ (Fock state) for plotting purposes; the choice of $\omega_a$ or $\omega_c$ alone does not matter.} 
\end{figure*}
\begin{figure*}[!htp]
\subfloat{\includegraphics[width=.32\textwidth,scale=1]{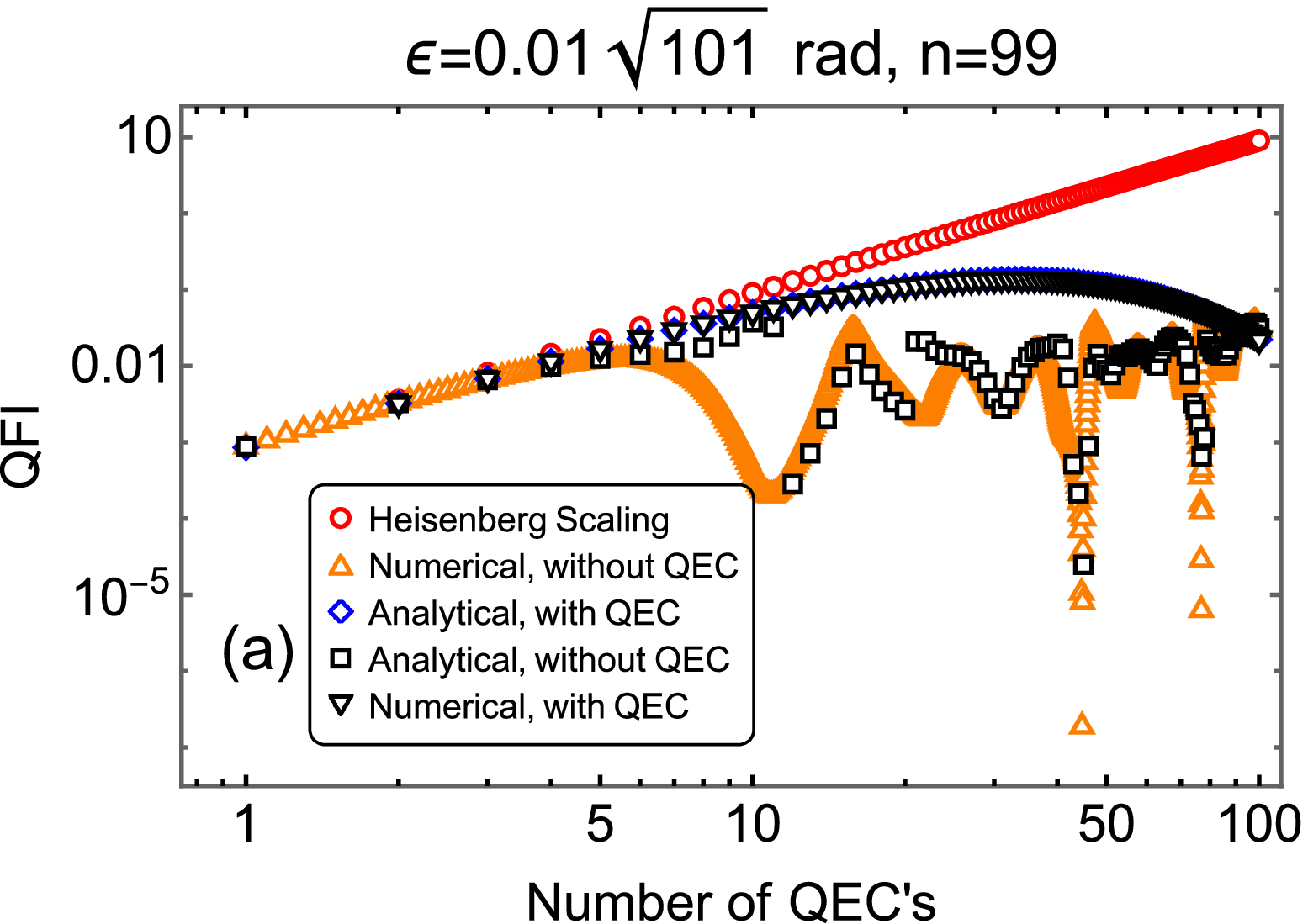}
\hfill}
\subfloat{
\includegraphics[width=.325\textwidth,scale=1]{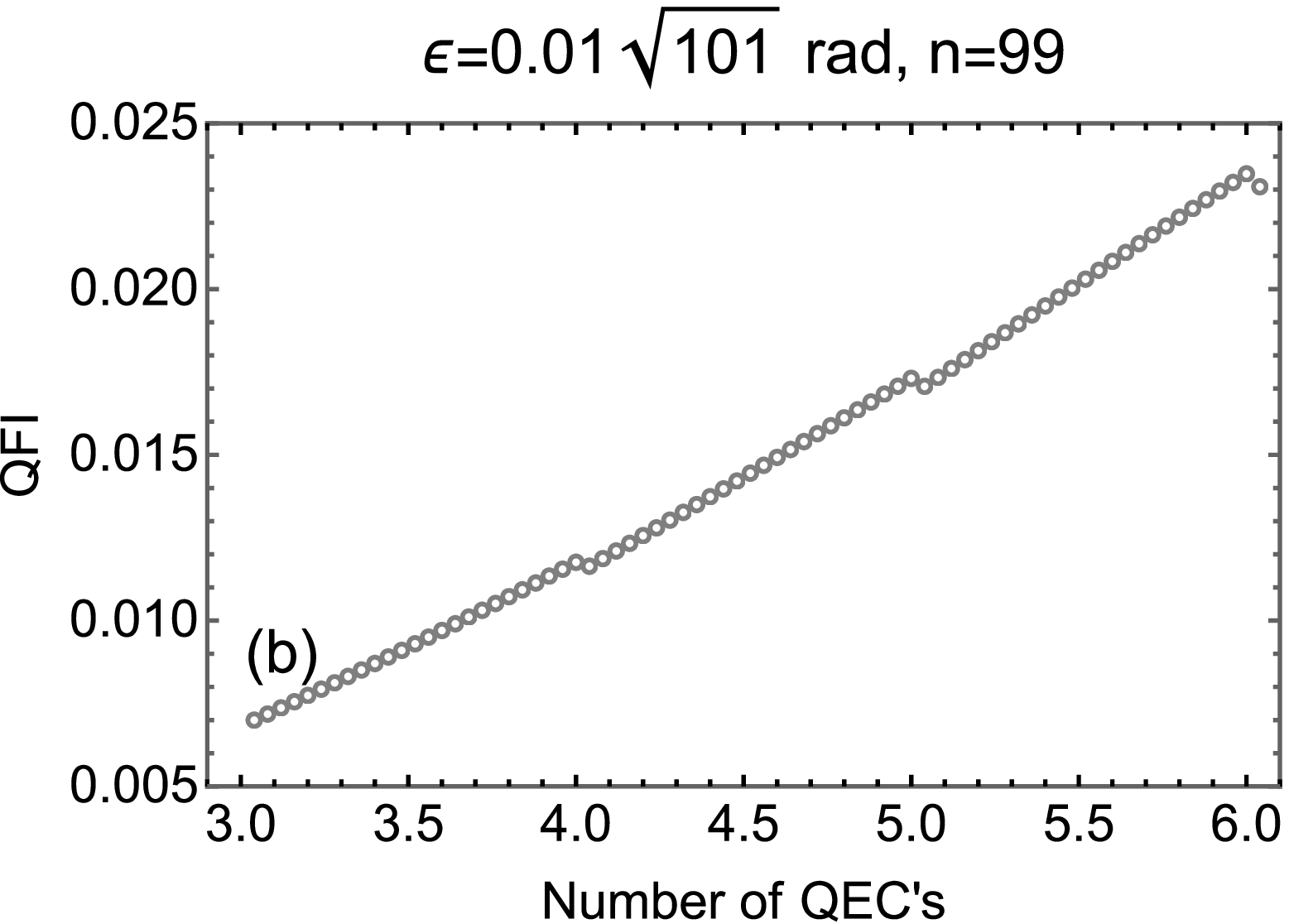}\hfill}
\subfloat{
\includegraphics[width=.325\textwidth,scale=1]{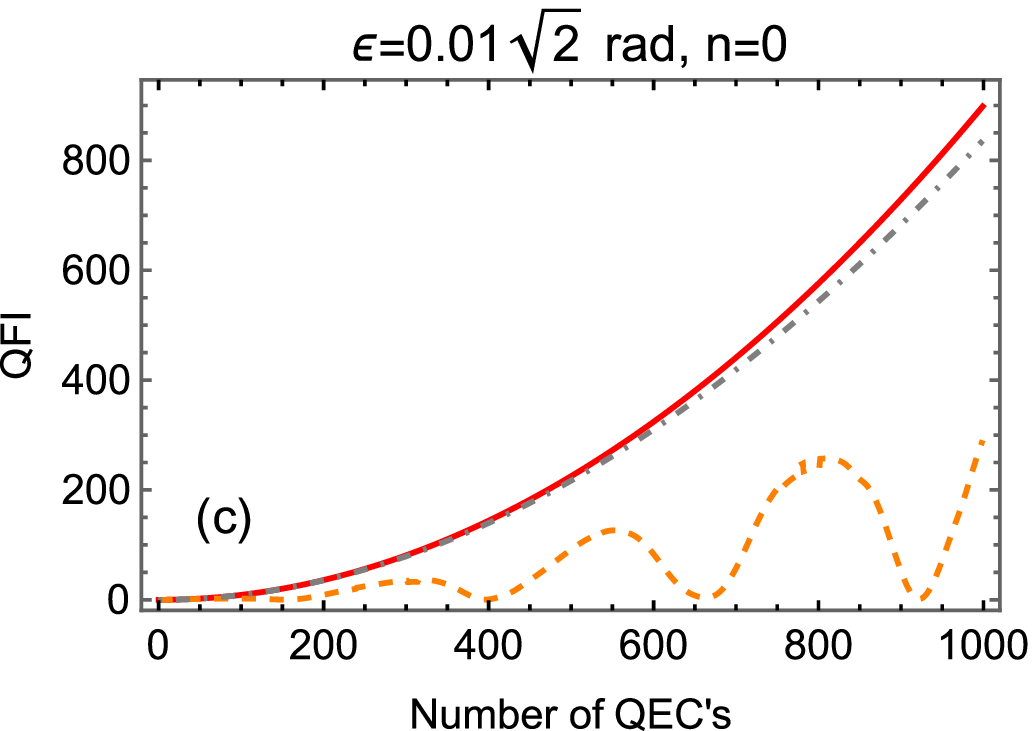}\hfill}

\caption{
\label{fig:epsart3} The QFI regarding the detuning frequency $\delta$ obtained at time $t$. Panel (a) shows a comparison between numerical and analytical results for QEC and non-QEC cases. Panel (b) shows a zoomed-in view for numerical results with QEC. Panel (c) shows the ideal Heisenberg scaling case $\mathcal{F}_{\delta}(t)=9t^2$(red curve) and the numerical simulation results for uncorrected (orange curve) and QEC (gray dash-dotted curve) cases. In all three plots, $\delta=2$ and $\Omega=2.$} 
\end{figure*}

The corrected density matrix $\rho^{\prime}{(\epsilon)}$ continues to evolve according to the same time evolution operator such that at time $t=\eta \epsilon$, when $\eta$ times of QECs have been applied for a fixed time interval $\epsilon$, the density matrix becomes $\rho_{S}^{\prime}(\eta \epsilon)$, where the nonzero entries are
\begin{equation}
\begin{aligned}
	{\rho_S}_{14}^\prime(\eta \epsilon)=\frac{1}{2}e^{i3\omega_c\eta \epsilon}\sum_m \bra{m}(U_\text{II}(\epsilon))_{11}^\eta \rho_E(0) (U_\text{II}^\dagger(\epsilon))_{44}^\eta	\ket{m},
\end{aligned}
\end{equation}
${\rho_S}_{41}^\prime(\eta \epsilon)={{\rho_S}_{14}^\prime}^\dagger(\eta \epsilon)$ and ${\rho_S}_{11}^\prime(\eta \epsilon)={\rho_S}_{44}^\prime(\eta \epsilon)=1/2$.
More generally, at time $t=\eta \epsilon + \tau$, the nonzero entries of the density matrix $\rho_{S}^{\prime}(\eta \epsilon+\tau)$ are
\begin{equation}
\begin{aligned}
	&{\rho_{S}}_{14}^{\prime}(\eta \epsilon+\tau)=\frac{1}{2} e^{i3(\omega_c\eta \epsilon+\tau)}\\&\times \sum_m \bra{m}U_\text{II}(\tau)_{11}(U_\text{II}(\epsilon))_{11}^\eta \rho_E(0) (U_\text{II}^\dagger(\epsilon))_{44}^\eta	(U_\text{II}^\dagger(\tau))_{44} \ket{m},
	\end{aligned}
\end{equation}
${\rho_{S}}_{41}^{\prime}(\eta \epsilon+\tau)={{\rho_{S}}_{14}^{\prime}}^*(\eta \epsilon+\tau)$ and ${\rho_{S}}_{ii}^{\prime}(\eta \epsilon+\tau)={\rho_S}_{ii}(\epsilon)$ for $i \in \{1,2,3,4\}$.
It is then straightforward to calculate QFI according to Eq.~(\ref{eq7}).

It has to be noted that, in the above discussions and following numerical simulations, we assume that the cavity remains at $\ket{n}$ even if multiple QECs have been applied. This is not exactly the case but should be a good approximation as long as $n$ is large enough and $t$ is not too large.

\subsection{Results}
Figure~\ref{fig:epsart2} shows the approximation results for $\delta$ measurement and Fig.~\ref{fig:epsart3} shows the results from numerical simulation for comparison purposes. To summarize, the QECs significantly improve the QFI gained, confirmed in both the numerical and the analytical cases.

In general, the shorter the period $\epsilon$ of QECs is, the better the effect is. If we define the error rate as the rate that at least one error occurs, $\varepsilon(\epsilon)=(1-4\rho_{11}(\epsilon))/\epsilon,$
then for the $s$-atoms case and under the limit $\epsilon \to 0$  we obtain
\begin{equation}\label{eq36}
	\varepsilon(\epsilon) = \frac{s}{4} (n+1) \Omega^2 \epsilon + \mathcal{O}(\epsilon^2),
\end{equation}
which means the error rate decreases linearly as $\epsilon$ decreases. We see this more evidently in Fig.\,\ref{fig:epsart4}(a) and Fig.\,\ref{fig:epsart4}(b). It is worth noting that in a noisy environment where $n$ is large, the correcting period $\epsilon$ has to be shortened accordingly for $\varepsilon \propto n\epsilon$ to remain constant.

\subsection{Discussion On the Generalized cases}
If we have $s$ atoms where $s>3$, the dimension of the Hilbert space will be $2^s$, which can be reduced to $s+1$ by the same aforementioned method. The corresponding $s$-qubit QEC is made such that any wrongly flipped qubits shall be unflipped according to the majority rule, assuming that the QEC is fast enough so that less than half of the qubits are wrongly flipped. At time $t=\eta \epsilon$, there are only up to four non-zero entries in  $\rho_{S}^{\prime}(\eta \epsilon)$, just as in the three-qubit case, which means that the only new terms we need to calculate are ${\rho_S}^\prime_{1,(s+1)}$ and ${\rho_S}^\prime_{(s+1),1}$.  At time $t=\eta \epsilon+\tau$, the general expression of $\rho_{S}^{\prime}(\eta \epsilon+\tau)$ for the $s$-atom case can be speculated:
\begin{equation}
\begin{aligned}
\rho_{ii}&=\frac{1}{2}[\left|u_{i 1}\right|^{2}+|u_{i, (s+1)}|^2]\\
&=\frac{1}{2}{s \choose i-1}  ([\chi(\tau)]^{i-1}[1-\chi(\tau)]^{s-i+1}\\
&+[\chi(\tau)]^{s-i+1}[1-\chi(\tau)]^{i-1}) \quad i \in\{1,2, \ldots, s+1\}
\end{aligned}
\end{equation}
where $\rho_{ii}=[\rho_{S}^{\prime}(\eta \epsilon+\tau)]_{ii}$ and $u_{ij} = (U_\text{II}(\tau))_{ij}$. All other entries are zeros except $\rho_{1,(s+1)}$ and $\rho_{(s+1),1}$. The results can be derived from calculations similar to those in Eq.\,(\ref{eq22}) and (\ref{eq25}), where the main difficulty lies in calculating $U_\text{II}(\epsilon)$, even with the large $n$ approximation applied. The above speculated formula is verified for $s=2,3$ and 4 cases.

\section{APPROXIMATE SOLUTION -- METHOD II}
\subsection{Derivation}
In the second method, we still consider the approximation $n \gg s$ such that $\hat{a}$ and $\hat{a}^\dagger$ can be treated as c-numbers $a$ and $a^*$, instead of operators for the moment; they will be restored to operators later. If we consider the $s$-atom case where $s$ is odd, then the original Hamiltonian [Eq.\,(\ref{eq1})] becomes
\begin{equation}
H=\sum_{i}^{s}\frac{1}{2}\left(\omega_{a} \sigma_{z}^{(i)}+\Omega a \sigma_{+}^{(i)}+\Omega a^* \sigma_{-}^{(i)}\right)+\omega_{c} a a^{*}.
\end{equation}
For each $i$, we have the eigenvalues $E_{\pm}=n \omega_{c} \pm \frac{1}{2} \Delta$ where $\Delta=\sqrt{n \Omega^2 + \omega_a^2}$ and $n=|a|^2$, and the eigenstates $\left|\Psi_{+}\right\rangle=c_{0}|0\rangle+c_{1}|1\rangle$ and 
$\left|\Psi_{-}\right\rangle=-c_{1}^{*}|0\rangle+c_{0}^{*}|1\rangle$
where $c_0=b/\sqrt{|b|^2+1}$, $c_1=c_0/b,$ and $b=(\Delta-\omega_a)/(\Omega a^*)$.
The eigenstates can be rearranged such that $
\ket{0}=c_0^* \ket{\Psi_+}-c_1\ket{\Psi_-}$ and 
$\ket{1}=c_1^* \ket{\Psi_+}+c_0\ket{\Psi_-}$
which after time $\epsilon$, becomes
\begin{equation}
\begin{aligned}
\ket{0}_\epsilon&=x_1(\epsilon)\ket{0}+x_2(\epsilon)\ket{1}\\
|1\rangle_{\epsilon} &=x_{3}(\epsilon)|0\rangle+x_{4}(\epsilon)|1\rangle,
\end{aligned}
\end{equation}
where $x_i(\epsilon)$ are some functions that are easily solvable.

If the initial state of the atoms is given by $
\left|\psi_{0}\right\rangle=\frac{1}{\sqrt{2}}(|0\rangle^{\otimes n}+\ket{1}^{\otimes n}),
$
then at time $\epsilon$ it becomes
\begin{equation}
\begin{aligned}
\left|\psi_{\epsilon}\right\rangle=\sum_{k=0}^{s}\left[\frac{1}{\sqrt{2}}(x_{1}^{k}(\epsilon) x_{2}^{s-k}(\epsilon)+x_{3}^{k}(\epsilon) x_{4}^{s-k}(\epsilon))\right]\\ \cdot \sum_{P} P\left(|0\rangle^{\otimes k}|1\rangle^{\otimes s-k}\right),
\end{aligned}
\end{equation}
where $P$ sums over all permutations on the $s$ atoms. We then apply QECs to unflip all wrongly flipped qubits according to the majority voting rule,
$
\rho(\epsilon) \stackrel{QEC}{\longrightarrow} \sum_k\hat{Q}_{k} \rho(\epsilon) \hat{Q}_{k}^{\dagger},
$
where
\begin{equation}
\begin{aligned}
\hat{Q}_k= P \left( \ket{0}^{\otimes s} \bra{0}^{\otimes k} \bra{1}^{\otimes (s-k)} + \ket{1}^ {\otimes s} \bra{0} ^{\otimes (s-k)}\bra{1}^{\otimes k}\right) \\
\end{aligned}
\end{equation}
for all $k=0,1,2,\cdots, (s-1)/2$. After one run of QEC, the density matrix at time $\epsilon$ becomes
\begin{equation}\label{eq52}
\begin{aligned} \rho^{\prime}(\epsilon) &=\sum_{k=\frac{s+1}{2}}^{s} {s \choose k}\left\{\left[\frac{1}{\sqrt{2}}\left(z_{0}(k, \epsilon)+z_{1}(k, \epsilon)\right)|0\rangle^{\otimes s}\right.\right.\\ &\left.\left.+\frac{1}{\sqrt{2}}\left(z_{0}(s-k, \epsilon)+z_{1}(s-k, \epsilon)\right)|1\rangle^{\otimes s}\right] \cdot \text{H.c} .\right\}\\
&= \sum_{k=\frac{s+1}{2}}^{s} {s \choose k}\hat{A}(k,\epsilon) \rho_0 \hat{A}^\dagger (k,\epsilon)\\
\end{aligned}
\end{equation}
where $z_0(k,t) = x_1^k(\epsilon) x_2^{s-k}(\epsilon)$, $z_1(k,t) = x_3^k(\epsilon) x_4^{s-k}(\epsilon)$ and
\begin{equation}
	\hat{A}(k,\epsilon)=\begin{pmatrix}
		z_0(k,\epsilon) & z_1(k,\epsilon)\\
		z_0(s-k,\epsilon) & z_1(s-k,\epsilon)
	\end{pmatrix}.
\end{equation}
which is written under the basis $\{\ket{0}^{\otimes s}, \ket{1}^{\otimes s}\}$.

We notice that $\hat{A}(k, \epsilon) \rho_{0} \hat{A}^{\dagger}(k, \epsilon)$ can be written as a sum of  $\hat{A}(k, \epsilon) \sigma_{\alpha} \hat{A}^{\dagger}(k, \epsilon)$ where $\alpha\in \{I,x,y,z\}$ (we denote $\sigma_I = 1$),
\begin{equation}\label{eq57}
\begin{aligned}
&\hat{A}(k, t) \sigma_\alpha \hat{A}^{\dagger}(k, t):=\left(\begin{array}{ll}a_{\alpha} & c_{\alpha} \\ c_{\alpha}^{*} & b_{\alpha}\end{array}\right)= \vec{v}_\alpha \cdot  \vec{\sigma}\\
\end{aligned}
\end{equation}
such that 
\begin{equation}
\hat{A}(k, \epsilon) \rho \hat{A}^{\dagger}(k, \epsilon)= 1/2\begin{pmatrix}
1,&x,&y,&z	
\end{pmatrix} V(k,\epsilon)  \vec{\sigma},
\end{equation}
where \begin{gather}
\vec{v}_\alpha=\begin{pmatrix}
	a_\alpha+b_\alpha,&c_\alpha+c_\alpha^*,&i(c_\alpha-c_\alpha^*),&a_\alpha-b_\alpha\nonumber
\end{pmatrix}^T\\
	\vec{\sigma}=(1,~\sigma_x,~\sigma_y,~\sigma_z)^T\\
	V(k, \epsilon)=1 / 2{s \choose k}\left(\begin{array}{llll}\vec{v}_{I}, & \vec{v}_{x},&\vec{v}_{y}, & \vec{v}_{z}\end{array}\right)\nonumber
\end{gather}
Here $a_\alpha, b_\alpha$, and $ c_\alpha$ are some terms that can be easily calculated.  
\begin{figure*}[!htp]

\subfloat{ \includegraphics[width=.32\textwidth ]{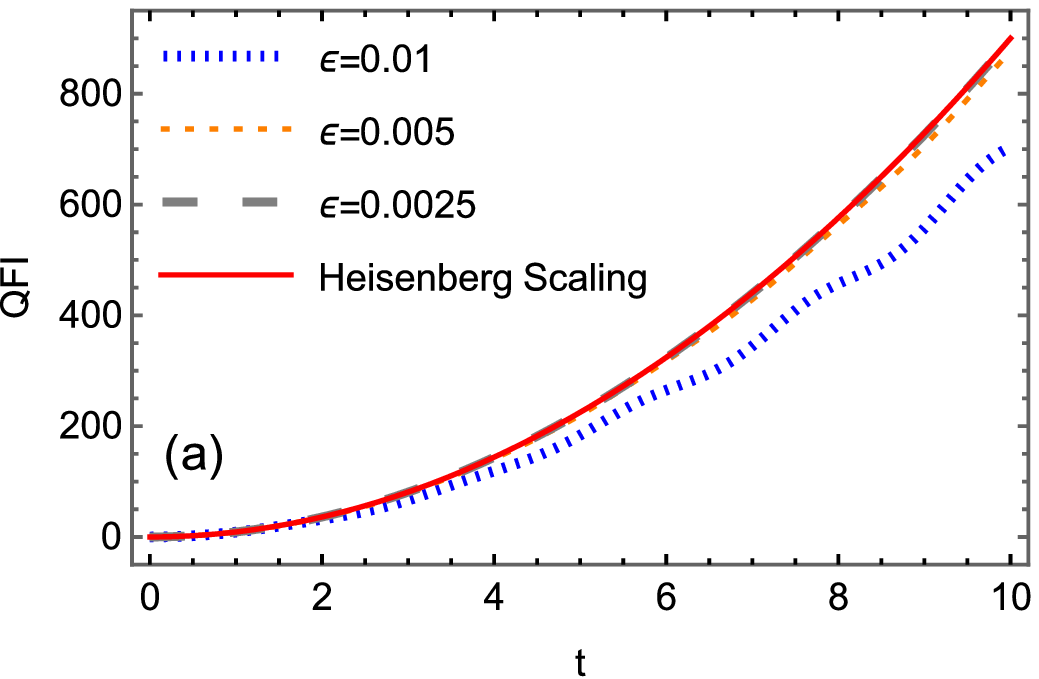}
\hfill}
\subfloat{
\includegraphics[width=.308\textwidth ]{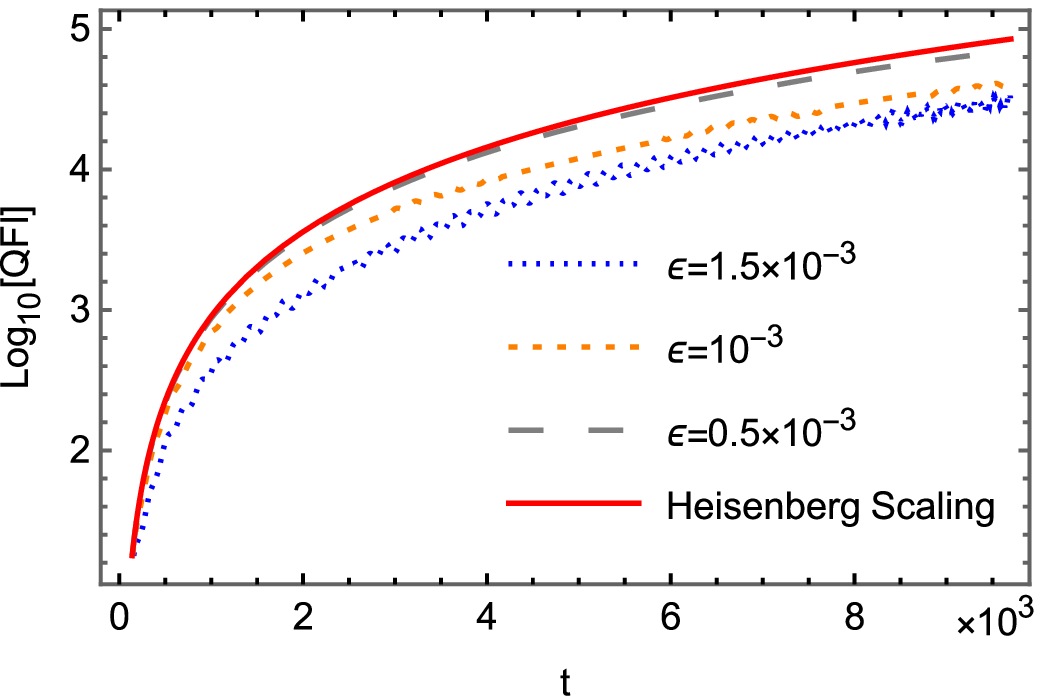}\hfill}
\subfloat{
\includegraphics[width=.337\textwidth]{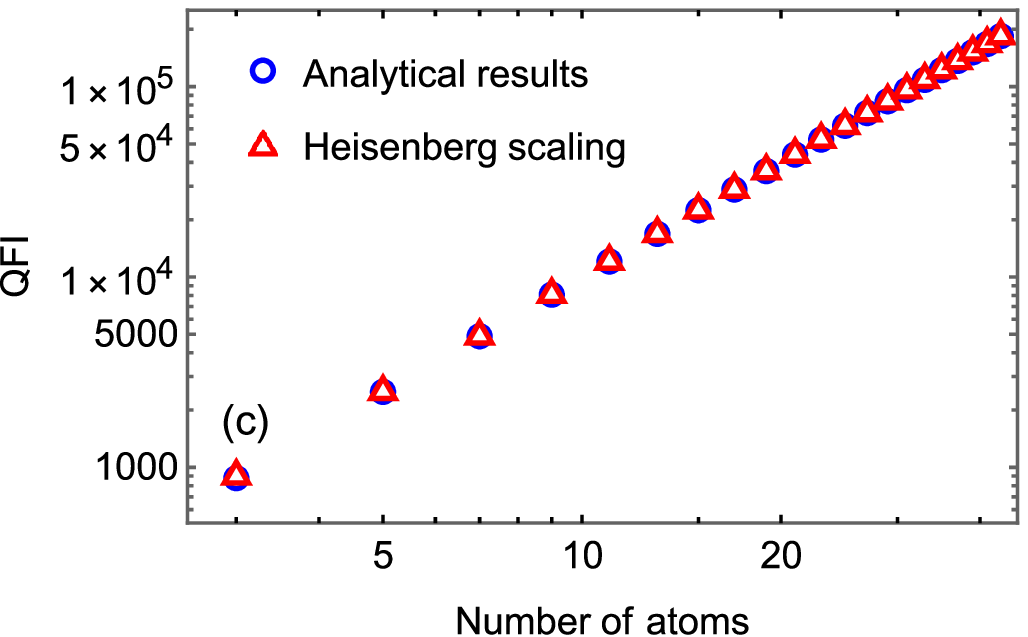}\hfill}

\caption{
\label{fig:epsart4} Panels (a) and (b) show the QFI with different QEC frequencies. The more frequent (smaller $\epsilon$) the QEC is, the better the QFI approaching to the Heisenberg scaling limit $\mathcal{F}_{\delta}(t)=9 t^{2}$ is (red curve). Panel (c) shows the QFI under different number of atoms, at $\epsilon=0.005$ and $t=10$. Note that different from Fig.~\ref{fig:epsart2} and \ref{fig:epsart3}, $\epsilon$ here is dimensional and has the same dimension as $t$. In all figures we take $\omega_c=2.5, \omega_a=0.5, n=100, \Omega=2 \text{ and } \rho_E=\ket{\alpha}\bra{\alpha}$ (coherent state) for plotting purposes.}
\end{figure*}
By Eq. ~(\ref{eq52}), the corrected density matrix can then be written as 
$
	\rho^\prime(\epsilon) =1/2(1,~ x, ~ y, ~ z) \mathcal{V}(\epsilon) \vec{\sigma},
$
where
$
	\mathcal{V}(\epsilon) = \sum_{k=(s+1)/2}^{s} V(k,\epsilon).
$
The benefit of the above approach is that to know the density matrix after $\eta$ number of QECs, we only need to raise the power of $\mathcal{V}(\epsilon)$ to $\eta$, which is very handy:
\begin{equation}
	\rho^\prime(\eta \epsilon) =1/2(1,~ x, ~ y, ~z) \mathcal{V}^\eta(\epsilon)\vec{\sigma}.
\end{equation}
It is worth noting that $\mathcal{V}(\epsilon)$ is always a $4\times4$ matrix, no matter how many atoms we have. For the three-atom case, $\mathcal{V}(\epsilon)$ turns out to be a block matrix $\mathcal{V}(\epsilon)=1_{1\times 1} \oplus V_{3\times3}$ with the analytic expressions of each entry explicitly obtainable, but too cumbersome to be shown here.

After obtaining $\rho^\prime(\eta \epsilon)$, the c-numbers $a$, $a^*$, and $n$ can be restored to operators. We can then calculate $\rho_S^\prime(\eta \epsilon)=\sum_m \bra{m}\rho^\prime (\eta \epsilon)\ket{m}$. It would be the best if we choose the the cavity to be initially at the coherent state $\ket{\alpha}$ such that
$
\hat{a}|\alpha\rangle=\alpha|\alpha\rangle
$
where $|\alpha|^2$ equals the mean number of photons. In this case, the c-numbers $a$ and $a^*$ can be replaced to $\sqrt{n}$  throughout the calculation, if considering the $n\gg s$ approximation such that 
$
\hat{a}^\dagger \ket{\alpha} \approx \alpha^* \ket{\alpha}.
$
\subsection{Results}
Figures \ref{fig:epsart4}(a) and \ref{fig:epsart4}(b) show QFI with different QEC frequencies, using the results from the second method. Smaller $\epsilon$ indicates more frequent QECs, which gives high QFI even in a prolonged time. Figure \ref{fig:epsart4}(c) showes that the Heisenberg scaling applies not only to the time $t$, but also the number of atoms $s$, \textit{i.e.} $\mathcal{F_\delta}(s) \propto s^2$. Note that different from Figs.~\ref{fig:epsart2} and \ref{fig:epsart3}, we assume the cavity is initially in the coherent state $\ket{\alpha}$, instead of the Fock state $\ket{n}$.

\section{Conclusion}
By using two different approaches, we show that by applying periodic QECs, we can achieve the Heisenberg scaling for an extended period of time on a three-qubit Tavis-Cummings model, where three two-level atoms interact with a single cavity mode, under the many photon approximation. Moreover, we show that the higher the frequency of the QEC, the longer the QFI can be kept at Heisenberg scaling, because the error rate decreases as the correcting frequency increases (Eq.~\ref{eq36}). Such Heisenberg scaling not only applies to the time $t$ (\textit{i.e.} QFI $\propto t^2$), but also applies to the number of atoms $s$ (\textit{i.e.} QFI $\propto s^2$)[Fig.~\ref{fig:epsart4}(c)].

\section{Acknowledgments.} We are grateful to Jing Yang for useful discussions. This research has been supported by the
US Army Research Office under Grant No. W911NF-18-10178 and the National Natural Science Foundation of China (NSFC) under Grant No. 12075323.

\appendix

\bibliographystyle{ieeetr}
\bibliography{apssamp}

\end{document}